\documentclass[10pt]{article}

\usepackage{amsthm}

\newcommand{\bb}[1]{\mbox{\boldmath ${#1}$}}

\newtheorem{theorem}{Theorem}
\newtheorem{lemma}[theorem]{Lemma}

\begin{document}

\title{ On deletion diagnostic statistic in regression }

\author{Myung Geun Kim\thanks{ Now a professor emeritus at Seowon University, Korea,~~ E-mail: mgkim@seowon.ac.kr}}

\date{}

\maketitle

\begin{abstract}
The change in the least squares estimator (LSE)  of a vector of regression coefficients due to a case deletion is often
used for investigating the influence of an observation on the LSE. A normalization of the change in the LSE using the
Moore-Penrose inverse of the covariance matrix of the change in the LSE is derived. This normalization turns out to be
a square of the internally studentized residual. It is shown that the numerator term of Cook's distance does not in
general have a chi-squared distribution except for a single case. An elaborate explanation about the inappropriateness
of the choice of a scaling matrix defining Cook's distance is given. By reflecting a distributional property of the
change in the LSE due to a case deletion,  a new diagnostic measure that is a scalar is suggested.  Three numerical
examples are given for illustration.
\end{abstract}

\vskip .3cm


\section{Introduction}

In regression, many diagnostic methods of identifying outliers or influential observations  have been suggested. Among
them, our interest will be confined to deletion diagnostic methods of investigating the influence of an observation on
the LSE of a vector of regression coefficients. The change in the LSE  of a vector of regression coefficients due to a
case deletion is a vector quantity and hence observations can not be ordered according to their influences. The changes
in the LSE due to case deletions are usually normalized or scaled so that observations can be ordered based on their
influences.
 Cook (1977) introduced a scaled distance by scaling the change in the LSE due to a case
deletion. However, the scaling matrix defining Cook's distance does not reflect a distributional property of the change
in the LSE due to a case deletion, which will usually lead to an incorrect detection of influential observations (see
Kim, 2017).

In Section 3.1, we will derive a normalization of the change in the LSE using the Moore-Penrose inverse of the
covariance matrix of the change in the LSE. This normalization turns out to be a square of the internally studentized
residual. In Section 3.2, we will show that the numerator term of Cook's distance does not in general have a
chi-squared distribution except for a single case. Furthermore we will give an elaborate explanation about the
inappropriateness of the choice of a scaling matrix defining Cook's distance. In Section 4 by reflecting a
distributional property of the change in the LSE due to a case deletion, we will suggest  a new diagnostic measure that
is a scalar. Hence observations are naturally ordered based on their influences provided by this diagnostic measure.
Three numerical examples are given for illustration.

\section{Preliminaries}

A linear regression model can be defined by
$$ {\bb y} = {\bb X}{\bb \beta} + {\bb \varepsilon},$$
where ${\bb y}$ is an $n \times 1$ vector of response variables, $ {\bb X}
 = ({\bb
x}_{1},...,{\bb x}_{n})^{T}$ is an $n \times p$ matrix of full column rank which consists of $n$ measurements on the
$p$ fixed independent variables,
  ${\bb \beta}$ is a $p \times 1$ vector of unknown regression coefficients, and
  ${\bb \varepsilon} = (\varepsilon_{1},...,\varepsilon_{n})^{T}$ is an $n \times 1$ vector of unobservable random errors in which
  $\varepsilon_{i}$ and $\varepsilon_{j}$ are uncorrelated for all $i$, $j$ ($i \neq j$).
Further it is assumed that each $\varepsilon_{i}$ has mean zero and  variance  $\sigma^2$.

The LSE  of ${\bb \beta}$  is $ \hat{{\bb \beta}}  = ({\bb X}^{T}{\bb X})^{-1}{\bb X}^{T}{\bb y}$ which is an unbiased
estimator of ${\bb \beta}$ and the covariance matrix of $ \hat{{\bb \beta}}$ is $\mbox{cov}(\hat{{\bb \beta}})=\sigma^2
({\bb X}^{T}{\bb X})^{-1}$. We write the hat matrix as ${\bb H} = (h_{ij}) = {\bb X}({\bb X}^{T}{\bb X})^{-1}{\bb
X}^{T}$. The residual vector  is ${\bb e} = (e_{1},...,e_{n})^{T} = ({\bb I}_n  - {\bb H}){\bb y}$, where ${\bb I}_n$
is the identity matrix of order $n$.  An unbiased estimator of $\sigma^{2}$ is  $ \hat{\sigma}^{2} = {\bb e}^{T}{\bb
e}/(n-p)$. More details can be found in Seber (1977).

\section{Deletion diagnostic measures}

The LSE of ${\bb \beta}$  computed without the $i$-th observation is written as $\hat{{\bb \beta}}_{(i)}$. Then we have
$$
\hat{{\bb \beta}} - \hat{{\bb \beta}}_{(i)} = ({\bb X}^{T}{\bb X})^{-1}{\bb
x}_{i}\frac{e_{i}}{1-h_{ii}}\hspace{.4cm}(i=1,...,n)
$$
whose derivation can be found in Miller (1974).

 The mean vector of $\hat{{\bb \beta}} - \hat{{\bb \beta}}_{(i)}$ is zero and its
covariance matrix is
$$
\mbox{cov}(\hat{{\bb \beta}} - \hat{{\bb \beta}}_{(i)})=\frac{\sigma^2}{1-h_{ii}}{\bb V}_{i},
$$
where
$$
{\bb V}_{i} =({\bb X}^{T}{\bb X})^{-1}{\bb x}_{i}{\bb x}_{i}^{T}({\bb X}^{T}{\bb X})^{-1}.
$$
The rank of ${\bb V}_{i}$ is one. It is easily shown that  ${\bb x}_i^T ({\bb X}^{T}{\bb X})^{-2} {\bb x}_i $ is the
only  nonzero eigenvalue of ${\bb V}_{i}$   and its associated eigenvector is $({\bb X}^T {\bb X})^{-1} {\bb x}_i$. For
more details, refer to Kim (2015).

The difference $\hat{{\bb \beta}} - \hat{{\bb \beta}}_{(i)}$ is used for investigating the influence of the $i$-th
observation usually through a normalizing or scaling process as follows
$$
(\hat{{\bb \beta}} - \hat{{\bb \beta}}_{(i)})^{T}{\bb M}(\hat{{\bb \beta}} - \hat{{\bb \beta}}_{(i)}),
$$
where ${\bb M}$ is an appropriately chosen matrix of order $p$. In the following two subsections we will discuss two
kinds of choices of ${\bb M}$.

\subsection{A normalization of $\hat{{\bb \beta}} - \hat{{\bb \beta}}_{(i)}$ using Moore-Penrose inverse}

The covariance matrix of $\hat{{\bb \beta}} - \hat{{\bb \beta}}_{(i)}$ is singular so that it is not invertible.
 Following the lines
given in the proof of  Theorem 5.1 of Schott (1997), we have the Moore-Penrose inverse of $V_i$ as
$$
{\bb V}_{i}^{+} = [{\bb x}_{i}^{T}({\bb X}^{T}{\bb X})^{-2}{\bb x}_i ]^{-2} {\bb V}_i .
$$
Hence the Moore-Penrose inverse of $\widehat{\mbox{cov}(\hat{{\bb \beta}} - \hat{{\bb \beta}}_{(i)})}$ is computed as
$$
[\widehat{\mbox{cov}(\hat{{\bb \beta}} - \hat{{\bb \beta}}_{(i)})}]^{+} =
\left(\frac{\hat{\sigma}^2}{1-h_{ii}}\right)^{-1} [{\bb x}_{i}^{T}({\bb X}^{T}{\bb X})^{-2}{\bb x}_i ]^{-2} {\bb V}_i .
$$
By using the Moore-Penrose inverse of $\widehat{\mbox{cov}(\hat{{\bb \beta}} - \hat{{\bb \beta}}_{(i)})}$, a normalized
distance between $\hat{{\bb \beta}}$ and $\hat{{\bb \beta}}_{(i)}$  can be obtained as
$$ (\hat{{\bb \beta}} -
\hat{{\bb \beta}}_{(i)})^{T}[\widehat{\mbox{cov}(\hat{{\bb \beta}} - \hat{{\bb \beta}}_{(i)})}]^{+}(\hat{{\bb \beta}} -
\hat{{\bb \beta}}_{(i)}) =\frac{e_{i}^{2}}{\hat{\sigma}^2 (1-h_{ii})}.
$$
 This normalization of $\hat{{\bb \beta}} - \hat{{\bb \beta}}_{(i)}$   is just a square
of the $i$-th internally studentized residual (see Eq. (4.6) of Chatterjee and Hadi, 1988 for the internally
studentized residuals).

\subsection{Cook's distance}

We will assume hereafter  that the error terms have a normal distribution with mean zero and variance $\sigma^2$. Based
on a confidence ellipsoid for ${\bb \beta}$, Cook (1977) introduced a diagnostic measure which can be expressed as
\begin{eqnarray*}
D_{i} &=& \frac{1}{p}(\hat{{\bb \beta}} - \hat{{\bb \beta}}_{(i)})^{T} [\widehat{\mbox{cov}(\hat{{\bb \beta}})}]^{-1} (\hat{{\bb \beta}} - \hat{{\bb \beta}}_{(i)})  \\
&=&\frac{1}{p\hat{\sigma}^{2}}(\hat{{\bb \beta}} - \hat{{\bb \beta}}_{(i)})^{T}({\bb X}^{T}{\bb X})(\hat{{\bb \beta}} - \hat{{\bb \beta}}_{(i)})  \\
&=& \frac{1}{p} \frac{h_{ii}}{1-h_{ii}} \frac{e_{i}^{2}}{\hat{\sigma}^2 (1-h_{ii})}.
\end{eqnarray*}
Cook's distance $D_{i}$ is a scaled distance between $\hat{{\bb \beta}}$ and $\hat{{\bb \beta}}_{(i)}$  using
$\widehat{\mbox{cov}(\hat{{\bb \beta}})}$.

\subsubsection{On comparing $D_{i}$  to the percentiles of the central $F$-distribution}
 The quantity
$$
\frac{1}{\sigma^{2}}(\hat{{\bb \beta}} - {\bb \beta})^{T}({\bb X}^{T}{\bb X})(\hat{{\bb \beta}} - {\bb \beta})
$$
has a chi-squared distribution with $p$ degrees of freedom. However, the quantity
$$
\frac{1}{\sigma^{2}}(\hat{{\bb \beta}} - \hat{{\bb \beta}}_{(i)})^{T}({\bb X}^{T}{\bb X})(\hat{{\bb \beta}} - \hat{{\bb
\beta}}_{(i)})
$$
does not in general have a chi-squared distribution except for a single case, which will be explained in this
subsection. To this end,   we will use Theorem 9.10 of Schott (1997) restated in the following lemma for easy
reference.

\begin{lemma}
Assume that  a random vector ${\bb x}$ is distributed as a  $p$-variate normal distribution $N_{p} ({\bb 0}, {\bb
\Omega} )$, where ${\bb \Omega}$ is positive semidefinite. Let ${\bb A}$ be a $p \times p$  symmetric matrix. Then a
quadratic form ${\bb x}^T {\bb A} {\bb x}$ has a chi-squared distribution with $r$ degrees of freedom if and only if
${\bb \Omega} {\bb A}{\bb \Omega}{\bb  A}{\bb  \Omega} = {\bb \Omega} {\bb A}{\bb  \Omega}$ and $\mbox{tr}({\bb A}{\bb
\Omega}) = r$.
\end{lemma}
Note that $(\hat{{\bb \beta}} - \hat{{\bb \beta}}_{(i)})/\sigma$ has a  $p$-variate normal distribution with zero mean
vector and covariance matrix ${\bb V}_{i}/(1-h_{ii})$. In Lemma 1, taking
$$
 {\bb \Omega }= \frac{1}{1-h_{ii}}{\bb V}_{i}~~~\mbox{and}~~~{\bb A}={\bb X}^T {\bb X},
$$
we have
$$
 {\bb \Omega} {\bb A}{\bb  \Omega}{\bb  A}{\bb  \Omega} = \frac{h_{ii}^2}{(1-h_{ii})^3} {\bb V}_i ~~~\mbox{and}~~~
{\bb \Omega}{\bb  A}{\bb  \Omega} = \frac{h_{ii}}{(1-h_{ii})^2} {\bb V}_i .
$$
Only when $h_{ii} = 1/2$, the first condition ${\bb \Omega}{\bb  A}{\bb  \Omega}{\bb  A}{\bb  \Omega} = {\bb
\Omega}{\bb  A}{\bb  \Omega}$ holds. Next, since
$$
\mbox{tr}({\bb A}{\bb  \Omega} ) = \frac{h_{ii}}{1-h_{ii}},
$$
we have $\mbox{tr}({\bb A}{\bb  \Omega} ) = k$ for $k=1,...,p$ when $h_{ii} = k / (1+k)$. Two conditions in Lemma 1 are
satisfied only when $h_{ii} = 1/2$. Thus we have the following theorem.

\begin{theorem} \label{thm:xx}
For each $i$, the quantity
$$
\frac{1}{\sigma^{2}}(\hat{{\bb \beta}} - \hat{{\bb \beta}}_{(i)})^{T}({\bb X}^{T}{\bb X})(\hat{{\bb \beta}} - \hat{{\bb
\beta}}_{(i)})
$$
 has a chi-squared distribution with one degree of freedom only when $h_{ii} = 1/2$.

\end{theorem}

 Cook (1977) suggests that each $D_{i}$ is compared to the percentiles of the central $F$-distribution
$F(p,n-p)$. Each $D_{i}$ does not strictly have an $F$-distribution (see p.120 of Chatterjee and Hadi, 1988).  Also,
Theorem \ref{thm:xx} shows that the use of $F$-distribution as a distributional form and the choice of numerator
degrees of freedom for $p \geq 2$ are inappropriate.

\subsubsection{On the choice of ${\bb X}^{T}{\bb X}$ as a scaling matrix}

First we will consider a distributional property of a random vector with a singular covariance matrix for easy
understanding in the next paragraph.

\begin{theorem} \label{thm:sing}
Assume that a random vector ${\bb x}$ is distributed as a  $p$-variate normal distribution $N_{p} ({\bb 0}, {\bb
\Omega} )$, where the rank of ${\bb \Omega}$ is $q$ with $1 \leq q < p$. Then ${\bb x}$ takes values in the column
space of ${\bb \Omega}$ with probability one.

\end{theorem}

\begin{proof}
 Let the spectral decomposition of ${\bb \Omega}$ be
$$ {\bb \Omega} = {\bb \Gamma} {\bb \Lambda} {\bb \Gamma}^{T},$$
where ${\bb \Gamma}$ is an orthogonal matrix with its $k$-th column ${\bb \gamma}_{k}$ ($k=1,\cdots , p$) and ${\bb
\Lambda}$ is a diagonal matrix $\mbox{diag}(\lambda_{1}, \cdots , \lambda_{q}, 0, \cdots , 0)$ with positive
eigenvalues $\lambda_{1}, \cdots , \lambda_{q}$ of ${\bb \Omega}$. The set $\{ {\bb \gamma}_{1}, \cdots , {\bb
\gamma}_{p} \}$ forms an orthonormal basis for the $p$-dimensional Euclidean space.

 Let $R({\bb \Omega})$ be the column space of ${\bb
\Omega}$ and $N({\bb \Omega})$ be the null space of ${\bb \Omega}$. The set $\{ {\bb \gamma}_{1}, \cdots , {\bb
\gamma}_{q} \}$ is an orthonormal basis for  $R({\bb \Omega})$ while the set $\{ {\bb \gamma}_{q+1}, \cdots , {\bb
\gamma}_{p} \}$ is an orthonormal basis for  $N({\bb \Omega})$. Since $R({\bb \Omega})$ is the orthogonal complement of
$N({\bb \Omega})$, we have
$$
 {\bb x} \in R({\bb \Omega}) ~~\mbox{ if and only if}~~~ {\bb \gamma}_{j}^{T} {\bb x} = 0 ~~\mbox{ for all}~ j=q+1 , \cdots
, p,
$$
which yields
$$
\{ {\bb x} \not \in R({\bb \Omega}) \}  ~~ = ~~ \cup_{j=q+1}^{p} \{   {\bb \gamma}_{j}^{T} {\bb x} \neq 0 \} .
$$
For each $j=q+1 , \cdots , p$, the mean of ${\bb \gamma}_{j}^{T} {\bb x}$ is zero and its variance is ${\bb
\gamma}_{j}^{T} {\bb \Omega}  {\bb \gamma}_{j} = 0$. Hence the probability that ${\bb \gamma}_{j}^{T} {\bb x}$ is equal
to $0$ is
$$
P (  {\bb \gamma}_{j}^{T} {\bb x} = 0 ) = 1.
$$
Thus it is easy to see that $P ({\bb x}  \in R({\bb \Omega}) ) = 1 $, that is, ${\bb x}$ takes values in the column
space of ${\bb \Omega}$ with probability one.
\end{proof}

We consider the spectral decomposition of ${\bb X}^{T}{\bb X}$ as
$${\bb X}^{T}{\bb X} = {\bb G}{\bb L}{\bb G}^T ,$$
 where ${\bb L} = \mbox{diag} (l_1,$ $ \cdots,$ $
l_p )$ is a $p \times p$ diagonal matrix consisting of the eigenvalues of ${\bb X}^{T}{\bb X}$, ${\bb G} = ({\bb g}_1 ,
\cdots , {\bb g}_p )$ is a $p \times p$ orthogonal matrix,   and ${\bb g}_k$ is the eigenvector of ${\bb X}^{T}{\bb X}$
associated with the eigenvalue $l_k$. Each $D_{i}$ can be expressed as
$$
D_i = \frac{1}{p\hat{\sigma}^{2}} \sum_{k=1}^p l_k [(\hat{{\bb \beta}} - \hat{{\bb \beta}}_{(i)})^T {\bb g}_k ]^2 . $$
The terms $l_k$ and $(\hat{{\bb \beta}} - \hat{{\bb \beta}}_{(i)})^T {\bb g}_k$  play a specific role in determining
the magnitude of $D_i$ for each $i$.

Since the rank of $\mbox{cov}(\hat{{\bb \beta}} - \hat{{\bb \beta}}_{(i)})$ is one,  Theorem \ref{thm:sing} shows that
$\hat{{\bb \beta}} - \hat{{\bb \beta}}_{(i)}$ is distributed entirely along the line generated by the eigenvector
$({\bb X}^T {\bb X})^{-1} {\bb x}_i$ of ${\bb V}_{i}$ which is one-dimensional subspace of the $p$-dimensional
Euclidean space. Since the eigenvectors of ${\bb X}^{T}{\bb X}$ are orthogonal to each other, all the eigenvectors of
${\bb X}^{T}{\bb X}$ or $p-1$ eigenvectors are not in general parallel to the line generated by  $({\bb X}^T {\bb
X})^{-1} {\bb x}_i$ in which the random vector $\hat{{\bb \beta}} - \hat{{\bb \beta}}_{(i)}$ takes values with
probability one. The distance $D_i$ inevitably includes the components $l_k [(\hat{{\bb \beta}} - \hat{{\bb
\beta}}_{(i)})^T {\bb g}_k ]^2 /p\hat{\sigma}^{2}$ (for all  the $k$'s or $p-1$ $k$'s) associated with the axes ${\bb
g}_k$ different from the axis determined by the eigenvector $({\bb X}^T {\bb X})^{-1} {\bb x}_i$. These components of
$D_i$ become a source of distorting the real influence of the $i$-th observation on $\hat{{\bb \beta}}$ because the
coordinates $(\hat{{\bb \beta}} - \hat{{\bb \beta}}_{(i)})^T {\bb g}_k$ with respect to the axes ${\bb g}_k$ different
from the axis determined by the eigenvector  $({\bb X}^T {\bb X})^{-1} {\bb x}_i$ are probabilistically meaningless.
Hence the adoption of ${\bb X}^T{\bb X}$ for scaling the distance between $\hat{{\bb \beta}}$ and $ \hat{{\bb
\beta}}_{(i)}$ is not reasonable and
 the Cook's distance measure  can not in general correctly identify influential observations.
 More details can be found in Kim (2017).

 \section{A new diagnostic measure}

 We note that the rank of $\mbox{cov}(\hat{{\bb
\beta}} - \hat{{\bb \beta}}_{(i)})$ is one and only $({\bb X}^T {\bb X})^{-1} {\bb x}_i$ is the eigenvector  of ${\bb
V}_{i}$ associated with a nonzero eigenvalue.
 The eigenvector  $({\bb X}^T {\bb X})^{-1} {\bb x}_i$
forms one axis in the $p$-dimensional Euclidean space.   Theorem \ref{thm:sing} implies that among $p$ coordinates of
$\hat{{\bb \beta}} - \hat{{\bb \beta}}_{(i)}$ in the $p$-dimensional Euclidean space, only the coordinate of $\hat{{\bb
\beta}} - \hat{{\bb \beta}}_{(i)}$ with respect to the axis $({\bb X}^T {\bb X})^{-1} {\bb x}_i$ is probabilistically
meaningful. This coordinate (or its absolute value) as a scalar represents  naturally the influence of the $i$-th
observation which the difference $\hat{{\bb \beta}} - \hat{{\bb \beta}}_{(i)}$ reflects in the $p$-dimensional
Euclidean space, and it is computed as
$$
K_i  =   \frac{e_i }{1-h_{ii}} ~ ||({\bb X}^{T}{\bb X})^{-1}{\bb x}_i ||,
$$
 where $||{\bb a} ||^2  = {\bb a}^{T}{\bb a}$ for a column vector ${\bb a}$.
Hence it is reasonable to use the quantity $K_{i}$ as a diagnostic measure to investigate the influence of the $i$-th
observation on $\hat{{\bb \beta}}$. The quantities $K_{1}, \cdots , K_{n}$ are natually ordered according to their
magnitudes.
 A relatively large absolute
value of $K_i$ implies that the $i$-th observation is potentially influential. The quantity $K_i$ is invariant under
orthogonal transformations of the rows of ${\bb X}$. However, it is not in general invariant under nonsingular
transformations. For a nonsingular transformation  ${\bb X}{\bb A}$ with a $p \times p$ nonsingular matrix {\bb A}, we
use $(e_i /(1-h_{ii})) ||{\bb A}^{-1}({\bb X}^{T}{\bb X})^{-1}{\bb x}_i ||$ instead of $K_i$.

\subsection{Hald data}

The regression model with the intercept term  is fitted to the Hald data set (Draper and Smith, 1981) which consists of
13 observations on a single response variable and four independent variables. For the Hald data, our discussion is
confined to observations 3 and 8.
 Cook's distances show that
 observation 8 is  the most influential  ($D_8 = 0.394$) and observation 3 is the next ($D_3 = 0.301$).
 However, the   $K_i$ values
 show that observation 3 is the most influential   ($K_3 = -76.197$) and observation 8 is the next
($K_8 = -25.168$). An analysis of the sources of the $D_i$ values for observations 3 and 8  shows that
  the  $D_8$ value  enlarges the  real influence of observation 8  on $\hat{{\bb \beta}}$,
while the  $D_3$ value reduces the real influence of observation 3 (Kim, 2017). Hence the $D_{3}$ value does not
identify observation 3 as the most influential one even though the $K_3$ value identifies observation 3 as the most
influential one, and the $D_{8}$ value  identifies observation 8 as the most influential one
 even though  observation 8  is not the most influential based on the $K_i$ values.

\subsection{Body fat data}

We fit the regression model with the intercept term   to the  body fat data set (Neter et al., 1996, p.261) which has
20 measurements on a single response variable and three independent variables. An analysis of the body fat data is
confined to observations 1 and 3. Based on the $D_i$ values, observation 3 is the most influential  ($D_3 = 0.299$) and
observation 1 is the next ($D_1 = 0.279$). However, for the $K_i$ values, observation 1 has the largest absolute value
($K_1 = -72.922$) and observation 3 has $K_3 = -37.466$, not the second largest absolute value. An investigation of the
sources of the $D_i$ values for observations 1 and 3 shows that the  $D_3$ value enlarges the real influence of
observation 3  on $\hat{{\bb \beta}}$, while the $D_1$ value reduces the real influence of observation 1 (Kim, 2017).
Hence the $D_{1}$ value does not identify observation 1 as the most influential one even though the $K_1$ value
identifies observation 1 as the most influential one,  and the $D_{3}$ value  identifies observation 3 as the most
influential one  even though observation 3 does not have the largest absolute value based on the $K_i$ values.

\subsection{Rat data}

The regression model with the intercept term  is fitted to the  rat data set (Cook, 1977) which consists of 19
measurements on a single response variable and three independent variables. For the rat data, we confine our discussion
to observation 3. The $D_i$ values show that observation 3 is the most influential  ($D_3 = 0.930$). For the $K_i$
values, observation 3 has the largest absolute value ($K_3 = 2.694$). Both diagnostic measures lead to the same
conclusion that observation 3 is the most influential.  The extent to which the  $D_3$ value reflects the real
influence of observation 3 on $\hat{{\bb \beta}}$ is very high (Kim, 2017). Hence  the $D_3$ value  gives the same
result as the $K_3$ value.

\end{document}